\documentclass[twocolumn,english,prl,superscriptaddress]{revtex4-2}
\usepackage[T1]{fontenc}
\usepackage[latin9]{inputenc}
\usepackage{amsmath}
\usepackage{amssymb}
\usepackage{graphicx}
\usepackage{babel}
\usepackage{mathtools} 
\usepackage{stmaryrd} 

\newcommand{\vertiii}[1]{{\left\vert\kern-0.25ex\left\vert\kern-0.25ex\left\vert #1 
        \right\vert\kern-0.25ex\right\vert\kern-0.25ex\right\vert}}

\graphicspath{{figures/}} 
\begin{document}
\title{Structural Landscapes in Geometrically Frustrated Smectics}
\author{Jingmin Xia}
\affiliation{Mathematical Institute, University of Oxford, Oxford, UK}
\author{Scott MacLachlan}
\affiliation{Department of Mathematics and Statistics, Memorial University of Newfoundland,
St. John\textquoteright s, NL, Canada}
\author{Timothy J. Atherton}
\affiliation{Department of Physics and Astronomy, Tufts University, Medford, MA 02155,
USA}
\author{Patrick E. Farrell}
\affiliation{Mathematical Institute, University of Oxford, Oxford, UK}
\begin{abstract}
A phenomenological free energy model is proposed to describe
the behavior of smectic liquid crystals, an intermediate phase that exhibits orientational order and layering at the molecular scale. Advantageous properties render the functional amenable to numerical simulation. The model is applied to a number of scenarios involving geometric frustration, leading to emergent structures such as focal conic domains and oily streaks and enabling detailed elucidation of the very rich energy landscapes that arise in these problems.
\end{abstract}
\maketitle
Smectic liquid crystals are complex fluids that exhibit orientational
order and a layered structure over macroscopic distances \citep{gennes-1972-article}. Since the
layers are nearly incompressible, an immediate consequence is that
the material prefers to locally adopt one of six families of surfaces
(in three dimensions) compatible with constant layer spacing
\citep{10.1051/jphys:019770038090110500,10.1103/physreve.61.1574,10.1080/02678290902814718,10.1080/02678299308036505}.
External constraints may force deformations of the smectic that are
incompatible with the layer constraint, leading to geometric frustrations
and the spontaneous assembly of a wide variety of textures with characteristic defect
structures of the smectic phase \citep{10.3406/bulmi.1910.3454,10.1080/02678290902814718}.
Driven by advances in surface control, there has been considerable
renewed interest in exploiting the ability of smectics to repeatably
self-assemble over device length-scales by using surface patterning
\citep{10.1080/02678290701618351,10.1021/la703717k}, topographical
features such as grooves \citep{10.1063/1.352978,10.1073/pnas.0407925101,10.1039/c2sm25416f,10.1039/d0sm01112f}
or posts \citep{10.1073/pnas.1214708109,10.1038/s41467-019-13012-9},
confinement in droplets \citep{10.1007/s00396-010-2367-7,10.1088/1361-648x/aa5706,10.1073/pnas.1115684109}
or curved surfaces more generally \citep{10.1007/s101890170134},
to produce emergent patterns \citep{10.1021/acs.langmuir.5b02508,10.1002/adom.201500153}
that are  optically active as lenses, gratings \citep{10.1103/physreve.70.011709},
photonic crystals \citep{10.1002/adom.201500153} or lithographic templates \citep{10.1002/adfm.201001303}.
Moreover, defect structures in the texture act to efficiently trap
dispersed micro- or nano-particles, making smectics useful for hierarchical
\citep{10.1080/02678292.2018.1502371,10.1002/adfm.201602729,10.1021/acs.nanolett.9b04347}
or synergistic \citep{10.1039/c5sm01458a} assembly processes that could potentially
be adopted for metamaterial, sensor or solar cell production.
Since many
of the remarkable properties of smectics arise because of the geometric
and topological consequences of layering, they form a paradigmatic
model system to understand geometric frustration in other lamellar
phases such as block copolymers \citep{10.1039/c2cs35115c,10.1039/c3sm52607k},
membranes and vesicles \citep{10.1039/c6sm02736a,10.1073/pnas.1213994110}. 

The very complicated structures that emerge in frustrated smectics
have, however, proven to be very challenging to model mathematically. While many of the
observed textures have been understood through elegant geometric approaches \citep{10.1002/adom.201500153,10.1021/acs.langmuir.5b02508,10.1021/acs.langmuir.7b03351,10.1039/c5sm01458a,10.1073/pnas.1214708109,10.1103/physrevlett.104.257802,10.1103/physrevx.3.041026,10.1007/s101890170134,10.1080/02678290902814718,10.1103/physreve.61.1574,10.1103/physreva.26.3037,10.1051/jphys:0197700380120151100},
or by perturbing from the nematic phase \citep{10.1103/physreve.92.062511,10.1038/ncomms15453,10.1103/physreve.77.061702},
to date there have been few successful efforts to use numerical methods
to predict the structures adopted by smectics in general configurations. Such methods could
be of great benefit to structure prediction where the defects cannot
be observed optically, for example in thin films \citep{10.1002/adma.201103791,10.1039/c5sm02241j,Nemitz:2016ef,10.1103/physreve.76.041702,10.1103/physrevlett.96.027803}.
Furthermore, scenarios where partial smectic order exists, such as during
the transition from the nematic to the smectic phase, may exhibit very
complicated pre-transitional structures \citep{10.1038/ncomms15453,10.1039/c2sm07207f,10.1103/physreve.77.061702,10.1063/1.321663}
and few studies have addressed the connection between pattern formation
and the peculiar critical behavior of liquid crystals at the nematic-smectic transition \citep{10.1073/pnas.2000849117}.
Dynamical phenomena, such as time-varying layer spacing \citep{10.1080/10587259908025403},
interactions between embedded particles \citep{10.1021/acs.langmuir.7b03351}
and the evolution of smectic films and bubbles \citep{10.1002/cphc.201301183,10.1209/0295-5075/100/16003,10.1039/c9sm01181a}
also present difficult problems that appear to require numerical modelling.

One major obstacle to successful modelling of smectics is the complicated
nature of the smectic order.
 In the original theory of de Gennes \citep{gennes-1972-article},
the smectic phase is characterized by a complex order parameter $\psi(\mathbf{r})=|\psi(\mathbf{r})|e^{i\phi(\mathbf{r})}$
that contains both the amplitude and phase of the density modulations.
It is a remarkably successful approach, providing a theory of the nematic-smectic-A
transition analogous to the Ginzburg--Landau theory of superconductivity. 
Nonetheless, it presents certain challenges, as reviewed in Pevnyi
et al.~\citep{Pevnyi:2014kw}.
The first issue is due to the topology of the complex order parameter $\psi$ itself: $\mathrm{Im}(\psi)$ does not contain physical information.
Second, this model is formed on a \emph{coarse-grained} basis, i.e., this energy does not represent the local free energy density on the length scale of the smectic layers themselves.
To amend these issues, Pevnyi et al.\ propose a theory
formulated in terms of a real-valued variation $\delta\rho(\mathbf{r})$ from the average density and a director field $\mathbf{n}(\mathbf{r})$, the local axis of average molecular alignment.
Using a real-valued density variation avoids many of the problems of alternative approaches such as
using double-valued complex order parameters \citep{Pevnyi:2014kw}.
Nonetheless, this theory as presented is not able to reproduce half-charge defects, due to the presence of director discontinuities in these defects \citep{ball-2017-article}, which cannot be characterized by a continuous vector field.
For example, around a $\pm{1}/{2}$ defect where $\mathbf{n}$
rotates by $\pm\pi$ degrees, a discontinuity line where $\mathbf{n}$
reverses sign must exist.
In fact, since $\mathbf{n}$ enters the model only through the tensor $N = \mathbf{n} \otimes \mathbf{n} = \mathbf{n}_{i}\mathbf{n}_{j}$,
Pevnyi et al.\ solve for $N$ instead in their implementation. This allows them to represent half-charge defects \cite{ball-2017-article}, but numerically enforcing that $N$ is a line field (i.e., of the form
$\mathbf{n} \otimes \mathbf{n}$ for some unit vector $\mathbf{n}$) in minimization is difficult \cite{borthagaray2020}.

In this Letter, we formulate a theory of smectics suitable for finite
element simulation and apply it to several partially understood problems involving the configuration
of smectics between antagonistic boundary conditions, i.e., those that
favor opposing orientations incompatible with the layer constraint.
We quantitatively study the transition from uniform layering to the formation
of defects \citep{10.1051/jphyscol:1975152}, examine the role of
imposed surface orientation on the configuration of focal conic domains \citep{10.1039/c2sm07207f}
and predict the structure of oily streaks that occur in very thin
smectic films \citep{10.1103/physrevlett.96.027803,10.1103/physreve.70.011709,Nemitz:2016ef}. 

We begin with Pevnyi et al.'s proposed energy \citep{Pevnyi:2014kw},
\begin{align}
F(\delta\rho,\mathbf{n}) & =\int_{\Omega}\left[\frac{a}{2}\left(\delta\rho\right)^{2}+\frac{b}{3}\left(\delta\rho\right)^{3}+\frac{c}{4}\left(\delta\rho\right)^{4}\right.\nonumber \\
 & \left.+B\left|\mathcal{D}^{2}\delta\rho+q^{2}\mathbf{n}\otimes\mathbf{n}\delta\rho\right|^{2}+\frac{K}{2}\left|\nabla\mathbf{n}\right|^{2}\right], \label{eq:Pevnyi}
\end{align}
which is to be extremized to obtain stationary solutions $\delta\rho$
and $\mathbf{n}$ subject to the pointwise constraint $\mathbf{n}\cdot\mathbf{n}=1$.
The first three terms in \eqref{eq:Pevnyi} with coefficients $a$,
$b$ and $c$ are a Landau--de Gennes expansion of the free energy
and set the preferred value of $\delta\rho$ in the uniform state,
$q$ is the wavenumber of the layering, $B$ is a nematic-smectic
coupling parameter, $\mathcal{D}^2$ denotes the Hessian operator, $K$
is the elastic constant, and $\Omega$ is the domain of integration.
The functional \eqref{eq:Pevnyi} can be derived from density-functional
theory (based on a molecular statistical description), analogous to earlier work on smectics \citep{linhananta91, poinewierski91}. 

Noticing the fact that (\ref{eq:Pevnyi}) depends only on elements of the dyad $\mathbf{n}_{i} \mathbf{n}_{j}$,
Ball \& Bedford \citep{ballbed-2015-article} proposed to modify \eqref{eq:Pevnyi} by replacing $\mathbf{n}_i \mathbf{n}_j$ by a uniaxial representation $(Q/s+I_d/d)_{ij}$, leading to
\begin{equation}
    \label{eq:bbmodel}
    \begin{aligned}
        F(\delta\rho,Q) &=
    \int_\Omega \left[\frac{a}{2}(\delta\rho)^2 + \frac{b}{3}(\delta\rho)^3 + \frac{c}{4}(\delta\rho)^4 \right.\\
                        & \left. + B \left|\mathcal{D}^2 \delta\rho+q^2\left(\frac{Q}{s}+ \frac{I_d}{d}\right)\delta\rho \right|^2
         + \frac{K}{2}|\nabla Q|^2
    \right].
    \end{aligned}
\end{equation}
Here, $s$ is the scalar order parameter, $I_d$ $(d \in \{2,3\})$ is the identity matrix, and $Q$ is a tensor-valued order parameter.
There is no longer any constraint imposed on the state variables.
They proved existence of minimizers of their modified model, but did not pursue any numerical analysis, or realize any implementation.
One can anticipate numerical difficulties caused by having $s$ on the denominator, as it is likely to be near zero for defect
structures of physical interest.

Inspired by the modification from Ball \& Bedford \citep{ballbed-2015-article},
we propose the following alternative energy functional:
\begin{align}
F(\delta\rho,Q) & =\int_{\Omega}\left[\frac{a}{2}\left(\delta\rho\right)^{2}+\frac{b}{3}\left(\delta\rho\right)^{3}+\frac{c}{4}\left(\delta\rho\right)^{4}\right.\nonumber \\
 & + B\left|\mathcal{D}^{2}\delta\rho+q^{2}\left(Q+\frac{I_{d}}{d}\right)\delta\rho\right|^{2}\nonumber\\
 & \left. +\frac{K}{2}\left|\nabla Q\right|^{2}+f_{n}(Q)\right],
 \label{eq:Functional}
\end{align}
where the nematic bulk energy density $f_n(Q)$ is $-l \left(\text{tr}(Q^2)\right) + l \left(\text{tr}(Q^2)\right)^2$ in two dimensions
and $-\frac{l}{2} \left(\text{tr}(Q^2)\right) - \frac{l}{3} \left(\text{tr}(Q^3)\right) + \frac{l}{2} \left(\text{tr}(Q^2)\right)^2$ in three dimensions.

We pause to contrast \eqref{eq:Functional}
with Ball \& Bedford's formulation \eqref{eq:bbmodel}.
In order to avoid possible numerical issues caused when $s\approx 0$, we instead weakly enforce $s=1$ by adding the nematic bulk term $f_n$.  The global minimizer of the nematic bulk energy $\int_\Omega f_n(Q)$ is known to be a uniaxial $Q$-tensor with scalar order parameter $s=1$ \citep[Proposition 15]{majumdar-2010-article}.  Thus, inclusion of this term both promotes the favorable scalar order parameter and a tendency towards a uniaxial expression for $Q$.

A substantial difficulty in obtaining the numerical solution of the minimization problem with \eqref{eq:Functional} arises from the presence of the Hessian term, which requires $\delta \rho \in
\mathcal{H}^2$ (i.e., square-integrable functions with square-integrable first and second derivatives). A \emph{conforming} discretization requires the use of $\mathcal{C}^1$-continuous elements (i.e., the approximation is continuous with continuous first derivatives).
Constructing these finite elements is quite involved in practice, especially in the three dimensional case.
We therefore turn to the use of nonconforming discretizations following the so-called $\mathcal{C}^0$ \emph{interior penalty} approach \citep{brenner-2011-book}.
Essentially, we use $\mathcal{C}^0$-conforming elements (i.e., continuous without necessarily continuous first derivatives) and penalize inter-element jumps in the first derivatives to weakly enforce $\mathcal{C}^1$-conformity.
To this end, we add a penalty term to the energy functional \eqref{eq:Functional}, leading to
\begin{equation}
    F_\gamma(\delta \rho, Q) := F(\delta \rho, Q) + \sum_{e\in \mathcal{E}_I} \int_{e} \frac{\gamma}{2h_{e}^3} \left( \llbracket \nabla \delta\rho \rrbracket \right)^2.
\end{equation}
Here, $\gamma$ is the penalty parameter (we fix $\gamma=1$ throughout this work), $\mathcal{E}_I$ is the set of interior facets (edges/faces) of a mesh, $h_{e}$ denotes the size of an edge/face $e$, and the jump operator of a vector $\nabla w$ on a facet $e$ of two adjacent cells, labelled $K_-$ and $K_+$, is defined to be $\llbracket \nabla w \rrbracket = (\nabla w)_-\cdot \nu_- + (\nabla w)_+ \cdot \nu_+$ with $\nu_-$ and $\nu_+$ denoting the restriction of the outward normal to $K_-$ and $K_+$, respectively. The numerical analysis of this discretization will be reported elsewhere.
Using a $\mathcal{C}^0$ interior penalty method has the advantages of both convenience and efficiency: the weak form is simple, with only minor modifications
from a conforming method, and fewer degrees of freedom are used than with a fully discontinuous method.

We now apply our discretization of \eqref{eq:Functional} to a class of problems that encompasses
commonly used techniques to induce self-organized structures in smectics.
The liquid crystal is confined between two substrates treated to promote
different preferred molecular orientations and must somehow interpolate
between them, but unlike a nematic liquid crystal that can achieve
this smoothly, a smectic may be prevented from doing so due to the
layer constraint. 

\begin{figure}
\includegraphics{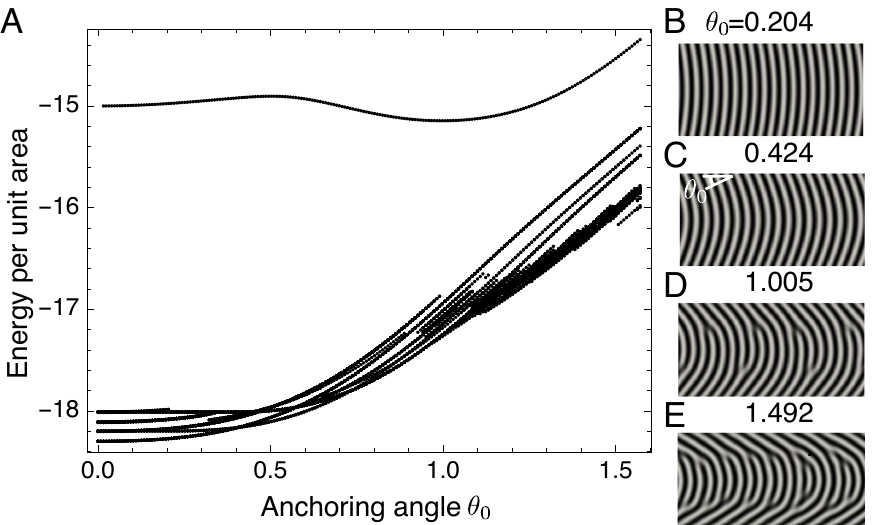}

\caption{\label{fig:WilliamsProblem}\textbf{Applying a bend deformation} to
a smectic liquid crystal. \textbf{A} Bifurcation diagram. \textbf{B-E}
Stable stationary solutions for different values of $\theta_0$. The visualization
displays the density variation $\delta \rho$.}
\end{figure}

As a simple example, proposed in the classic work of Williams \&
Kl\'eman \citep{10.1051/jphyscol:1975152}, consider the situation depicted
in Fig.~\ref{fig:WilliamsProblem} where we impose the director $\mathbf{n}_{e}=(\cos\theta_0,-\sin\theta_0)$ for fixed $\theta_0\in [0,\pi/2]$
at the lower boundary and $\mathbf{n}_{e}=(\ensuremath{\cos\theta_0,\sin\theta_0)}$
at the upper boundary.
The corresponding boundary data for the $Q$-tensor derived from $\mathbf{n}_e$ is given in the Supplemental Material \cite{supplemental}.
For $\theta_0=0$, the boundary conditions become
identical and the resulting configuration is with the layers extending
vertically between the substrates in the ``bookshelf'' geometry.
As $\theta_0$ is increased from zero, the boundary conditions impose
a bend deformation on the smectic. This can be accommodated in several
ways: by distributing the deformation over the vertical direction
(Fig.~\ref{fig:WilliamsProblem}B); by localizing the bend to a region
in the center with the layers flat and tilted in opposite directions
in the top and bottom of the domain (Fig.~\ref{fig:WilliamsProblem}C);
or by introducing edge disclinations to relieve the cost of elastic
deformation (Fig.~\ref{fig:WilliamsProblem}D,E).

The equilibrium structure as a function of $\theta_0$ is hence determined
by an energetic competition between the cost of bending and the cost
of introducing disclinations.
Using a technique called \emph{deflation} \citep{farrell-birkisson-2015-article},
we can compute a bifurcation diagram for this scenario and quantitatively
determine which of these solutions is the ground state as a function
of $\theta_0$ (Fig.~\ref{fig:WilliamsProblem}A).
Readers may refer to the Supplemental Material \citep{supplemental} for full details of the problem setup, an extended presentation of more stationary configurations computed in this scenario, and a video illustrating the lowest energy solutions found as $\theta_0$ is varied (all of which are stable).

\begin{figure}
\includegraphics[width=1\columnwidth]{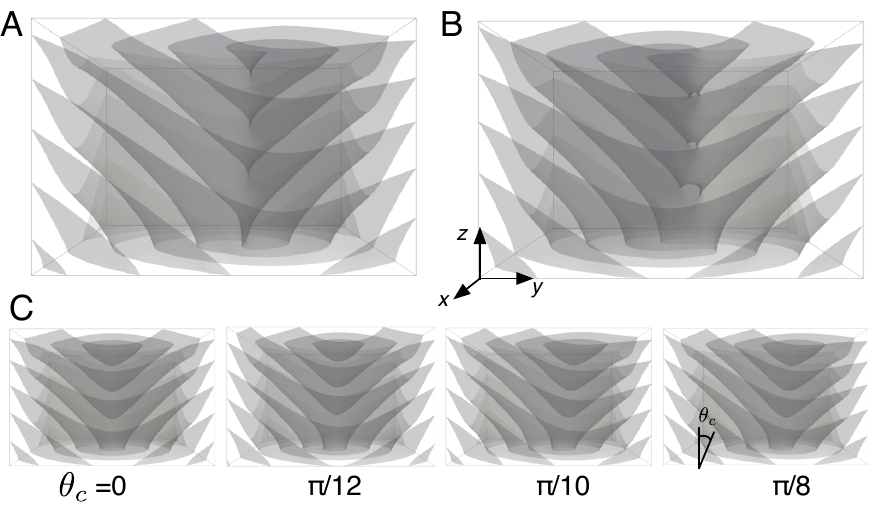}

\caption{\label{fig:FCD}\textbf{Focal conic domains} under tilted boundary
    conditions. 
     Example solutions with \textbf{A} single and \textbf{B} double screw dislocation defect at $\theta_c=\pi/12$. \textbf{C} Stable stationary solutions for different values of $\theta_c$.
Here, zero isosurfaces of the density variation $\delta \rho$ are displayed to visualize the layer structure of the smectic.
Among the three solutions shown for $\theta_c=\pi/12$, the FCD solution possesses the lowest energy value, while double screw dislocation solution has highest value.
}
\end{figure}

A more extreme scenario is where the preferred alignment axes at each
surface are perpendicular: one favors planar and the other vertical
alignment. The experimentally observed configurations in this case
are known as \emph{toroidal focal conic domains} (TFCDs): the smectic layers
adopt a configuration consisting of stacked interior sections of tori,
with a central line defect extending between the two substrates. TFCDs
may exist as isolated domains in a background of vertically oriented
smectic layers, or may self-assemble into a hexagonal lattice \citep{10.1038/ncomms15453,10.1039/c2sm07207f,10.1103/physreve.77.061702,10.1063/1.321663}.
If one of
the boundary conditions is perturbed, such as by introducing a small
preferred tilt at either substrate, asymmetric FCDs may arise where
the layers form from sections of Dupin cyclides \citep{10.1038/ncomms15453,10.1039/c2sm07207f}.

Despite the centrality of TFCDs in the study of smectics, and for
applications, prior numerical work on them has been limited to finding
solutions using modifications of the nematic theory \citep{10.1103/physrevlett.96.027803}. We
therefore verify that FCDs are stationary solutions of our functional (\ref{eq:Functional}) and characterize their response to tilted
boundary conditions.
Specifically, we perturb the zenith angle $\theta_c$ between the director and the $z$-axis in the boundary configuration (see detailed descriptions in the Supplemental Material \citep{supplemental}).
Displayed in Fig.~\ref{fig:FCD} is a sequence
of solutions as a function of $\theta_c$, the preferred tilt away from
the vertical at the upper substrate. All solutions displayed are stable.

As can be seen in Fig.~\ref{fig:FCD}C, we recover the
cylindrically symmetric TFCD for $\theta_c=0$; as $\theta_c>0$ the solution
becomes asymmetric (Fig.~\ref{fig:FCD}C) and the central defect line becomes a hyperbola,
as expected from geometry \citep{10.1007/s101890170134, 10.1080/02678290902814718, 10.1103/physreve.61.1574}. 
Three examples of the solution structures (including the single screw dislocation defect presented in Fig.~\ref{fig:FCD}B) at $\theta_c=\pi/12$ are animated in the Supplemental Material \citep{supplemental}.

\begin{figure*}
\includegraphics{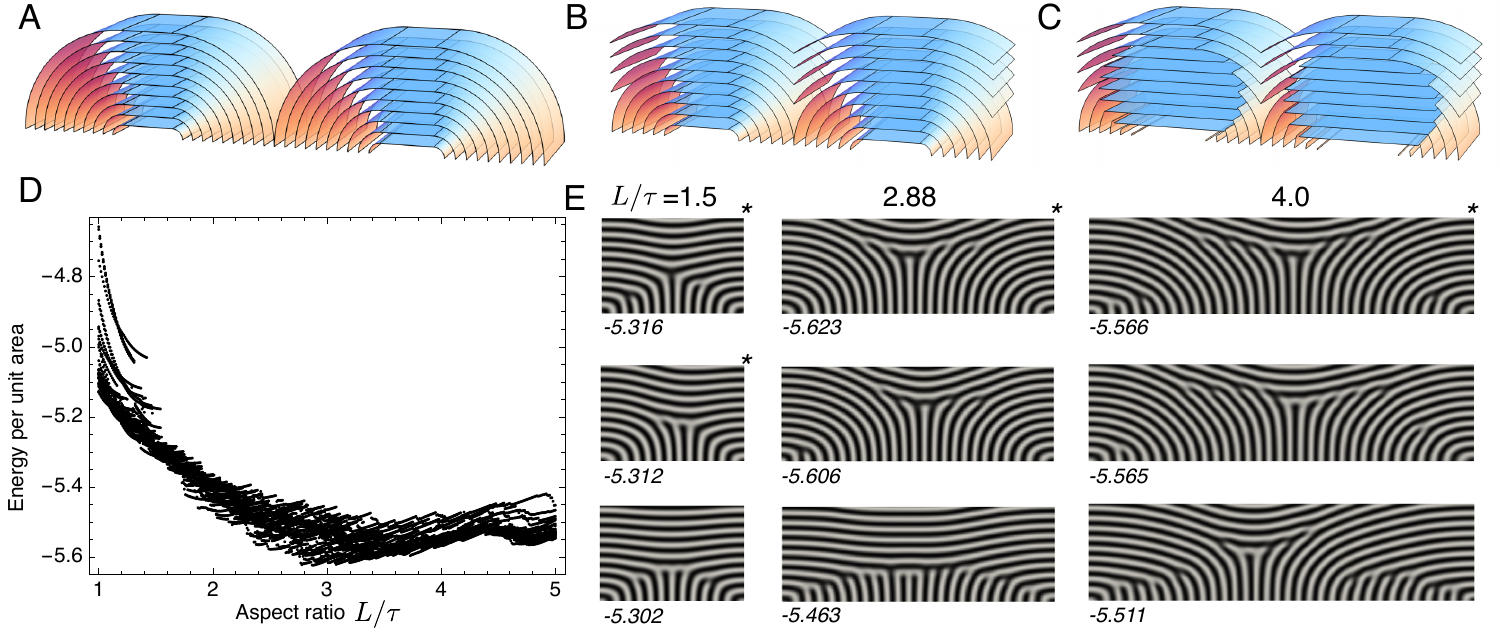}

\caption{\textbf{\label{fig:Oily-Streaks.-A-C}Oily streaks.} \textbf{A-C} Candidate
    structures proposed in Michel et al.~\citep{10.1103/physreve.76.041702} consistent with X-ray diffraction.
    \textbf{D} Bifurcation diagram of structures as a function of aspect ratio $L/\tau$.
\textbf{E} Selected stationary states obtained at different aspect ratios $L/\tau$. The top row represents the lowest energy solution found.
For each solution, the value of the energy functional per unit area is displayed below it with asterisks indicating stable profiles.}

\end{figure*}

For sufficiently thin films, the elastic energy cost of two dimensional
curvature of the layers observed in the FCD solutions becomes prohibitive.
Instead, the smectic adopts a configuration referred to as an ``oily streak'' texture \citep{10.1103/physreve.70.011709,10.1103/physrevlett.96.027803,Nemitz:2016ef}.
The structure is periodic in one direction parallel to the substrate
and spatially uniform in the other tangential direction; the periodicity
$L$ is experimentally found to increase linearly with the film thickness
$\tau$ such that $L\approx L_{0}+4.5\tau$ \citep{10.1103/physreve.76.041702}.
Addition of chiral dopants can be used to control the orientation
of the streaks \citep{Nemitz:2016ef}. 

X-ray diffraction experiments for films $0.15\mu m\le \tau\le0.35\mu m$
($47-110$ layers) indicate that the smectic
layer normals are continuously and uniformly distributed in orientation
with a significant additional peak for smectic layers that are parallel
to the plane of the substrate. An approximate layer structure proposed
by Michel et al.~\citep{10.1103/physreve.76.041702} consistent
with this data comprises periodic units incorporating sections of
cylinders joined to planes oriented parallel to the substrate (Fig.~\ref{fig:Oily-Streaks.-A-C}A). This structure implies, however, significant
deformations of the free interface with singular points between units;
while undulations of the smectic-air interface are observed by atomic force microscopy,
the amplitude is only around $1/5$ of the film thickness once the
finite size of the tip is accounted for. To address this, the same
authors consider more complex structures incorporating curvature walls \citep{10.1103/physreve.70.011709,10.1103/physrevlett.96.027803}
(Fig.~\ref{fig:Oily-Streaks.-A-C}B) that necessarily imply local
dilation of the layers or local melting into the nematic phase along
walls between units. 

For even thinner films $\tau\sim70nm$ (approx.~$22$ layers), X-ray diffraction
reveals an apparent excess of the planar region that cannot be explained
by either structure discussed so far \citep{10.1103/physrevlett.96.027803};
a possible structure that does so, and is consistent with the X-ray
data, is depicted in Fig.~\ref{fig:Oily-Streaks.-A-C}C and incorporates
an approximately hemicylindrical Rotating Grain Boundary (RGB) that
partitions the cylindrical component from the planar component. Such
a structure, with abruptly discontinuous layers, is energetically
very costly and was envisioned in \citep{10.1103/physrevlett.96.027803}
as a mesoscopic approximation: at the nanoscopic level, the RGB might
contain a network of dislocations to include the additional layers,
or locally melt into the nematic phase along the RGB. 

Hence, while these \emph{ansatz} models are very helpful in that they
provide an overall understanding of the structure and facilitate interpretation
of the experimental data, they incorporate coarse-grained features
such as the RGBs and, moreover, they are not calculated stationary
states of an appropriate free energy functional. Understanding the detailed structure of the oily
streaks therefore remains an important open problem. 

We again use the deflation technique to explore the stationary states of (\ref{eq:Functional})
on a rectangular domain of fixed vertical dimension and varying aspect ratio $L/\tau$.
For simplicity we do not allow for variation of the free surface,
which will be an area of future work, but instead impose weak anchoring
conditions.
As with the other numerical experiments, full details of the boundary conditions, solvers and choices of parameters are documented in the Supplemental Material \citep{supplemental}.
Furthermore, a video showing the lowest energy configurations as the aspect ratio $L/\tau$ varies is also included, all of which are stable.

A partially enumerated energy landscape is displayed in Fig.~\ref{fig:Oily-Streaks.-A-C}D, showing an extremely dense thicket of solutions, but qualitatively supporting earlier work in that an overall minimizer occurs at an aspect ratio of around $3$, which is similar to experimental values even with no parameter tuning performed here. 
Close examination of the energy
landscape, together with the corresponding solution set, shows many small discontinuous jumps that result from delicate
commensurability effects, whereby certain sizes of domain are compatible
with a given periodicity of the layers as well as from variations in the number of defects and their detailed placement. Similar effects have been
observed when other periodic liquid crystals such as cholesterics
are confined in domains that promote geometric frustration~\citep{emerson2017}.

The solution set obtained contains examples reminiscent of previously proposed structures (Fig.~\ref{fig:Oily-Streaks.-A-C}E). The minimum energy states found at different aspect ratio contain cylindrical sections mediated by a defect-filled region reminiscent of the mesoscopic rotating grain boundaries. Other solutions displayed in the lowest row of Fig.~\ref{fig:Oily-Streaks.-A-C}E are quite different from those heretofore proposed, where regions of relatively vertically
oriented layers sit atop cylindrical regions interspersed with defects.
Each of these incorporates a greater proportion of vertical layers
relative to the hemicylindrical-planar \emph{ansatz} of Fig.~\ref{fig:Oily-Streaks.-A-C}A,B
and may provide alternative structures for oily
streaks in ultrathin films. In future work, the boundary conditions at the top interface should be carefully reconsidered, including the incorporation of a free interface. 

\emph{Conclusion}---We have formulated a free energy functional for smectics that is amenable to finite element simulation, and applied it to scenarios involving boundary conditions that are incompatible with uniform smectic order; our new model successfully reproduces, even without careful tuning of parameters, a number of experimentally observed and theoretically expected phenomena, as well as producing new candidate structures for thin smectic films that are explicitly stationary states of an energy functional. We also demonstrate how to overcome a less obvious difficulty with numerical studies of smectics and layered media generally: the solution landscapes are extremely dense due to the presence of defects. The combination of our model together with the deflation technique enables detailed exploration of this landscape, enabling us to isolate both the ground state and low-lying excited states that may be observed in physical systems. 

\begin{acknowledgments}
The work of JX is supported by the National University of Defense Technology and the EPSRC Centre for Doctoral Training in Partial Differential Equations [grant number EP/L015811/1].
The work of SM was partially supported by an NSERC Discovery Grant.
The work of PEF was supported by EPSRC grants EP/R029423/1 and EP/V001493/1.
The work of TJA was partially supported by NSF grants DMR-1654283 and OAC-2003820. 
\end{acknowledgments}

\bibliographystyle{apsrev4-2}
\bibliography{smectics}

\end{document}


\title{Supplemental Materials: Structural Landscapes in Geometrically Frustrated Smectics}
\author{Jingmin Xia}
\affiliation{Mathematical Institute, University of Oxford, Oxford, UK}
\author{Scott MacLachlan}
\affiliation{Department of Mathematics and Statistics, Memorial University of Newfoundland,
St. John\textquoteright s, NL, Canada}
\author{Timothy J. Atherton}
\affiliation{Department of Physics and Astronomy, Tufts University, Medford, MA,
USA}
\author{Patrick E. Farrell}
\affiliation{Mathematical Institute, University of Oxford, Oxford, UK}

\maketitle

\section{Numerical Simulations}

We consider three examples: the defect-free example from the work of
Williams \& Kl\'eman \cite{10.1051/jphyscol:1975152}, a focal conic
domain simulation, and an oily streaks simulation. Throughout all
three, we use $\mathcal{C}^0$-continuous finite-element
pairs for $(\delta\rho,Q)$ with the interior penalty term described in the main text.  In two dimensions, we use quadrilateral meshes where the space $\mathbb{CG}_k$ is given by tensor products of polynomials of up to degree $k$ in each coordinate direction, yielding the spaces of piecewise biquadratic functions for $\mathbb{CG}_2$ and piecewise bicubic functions for $\mathbb{CG}_3$.  We restrict $Q$ to be a symmetric and traceless tensor, so it has two independent components in two dimensions. We thus seek the components of $Q$ in $\mathbb{CG}_2^2$ and $\delta\rho$ in $\mathbb{CG}_3$.  We use hexahedral meshes in three dimensions with similar tensor-product spaces, yielding the space of piecewise triquadratic functions for $\mathbb{CG}_2$ and piecewise tricubic functions for $\mathbb{CG}_3$.  In three dimensions, $Q$ has five independent components, so we seek its components in $\mathbb{CG}_2^5$, while retaining $\delta\rho$ in $\mathbb{CG}_3$.

Since the PDE problem to be solved is nonlinear, we use Newton's method with $L^2$ linesearch \cite[Algorithm 2]{brune2015} as the outer nonlinear solver.
The nonlinear solve is deemed to have converged when the Euclidean norm of the residual falls below $10^{-8}$, or reduces from its initial value by a factor
of $10^{-8}$, whichever comes first.
For the inner solves, the linearized systems are solved using the sparse LU factorization library MUMPS \cite{mumps}.
The solver described above is implemented in the Firedrake \cite{firedrake} library, which relies on PETSc \cite{petsc} for solving the resulting linear systems.
The mesh scale, $h_e$, employed in the $\mathcal{C}^0$ interior penalty approach is chosen to be the average of the diameters of the cells on either side of an edge/face.

\subsection{The deflated continuation algorithm}

The so-called \emph{deflation continuation} algorithm is used to find multiple solutions throughout all three of the scenarios considered.
More details can be found in \cite{farrell-birkisson-2015-article}.
We give a self-contained explanation of the algorithm here.

Consider a general parameter-dependent nonlinear problem 
\begin{equation}\label{problem}
    f(u,\lambda) = 0\quad \text{for }u\in U\ \text{and } \lambda\in \mathbb{R},
\end{equation}
where $U$ is an admissible space for $u$ and $\lambda$ is the parameter. In our
context $f$ is the residual of the Euler--Lagrange equation for the smectic energy functional, $u$ represents the pair $(\delta \rho, Q)$, and
$\lambda$ represents the director angle $\theta_0$ imposed on the boundary (Scenario I),
the angle $\theta_c$ between the director and the $z$-axis on the top surface (Scenario II),
or the aspect ratio $L/\tau$ of the domain (Scenario III).

For a given $\lambda$ there may be multiple solutions $u$ that satisfy this equation,
corresponding to different experimental outcomes for the same
experimental setup (due to multistability, i.e., the
existence of distinct local minimizers).
Local maxima and saddle points of the smectic energy functional will also be solutions of \eqref{problem}.
For a fixed parameter $\lambda^{\star}$, problem \eqref{problem} then becomes
\begin{equation}\label{fix-problem}
    G(u)\coloneqq f(u,\lambda^{\star}) = 0.
\end{equation}
We first apply the classical Newton iteration from an initial guess $u^0$ to find a solution $u^{\star}$ of \eqref{fix-problem}. We then \emph{deflate} this solution. The goal of deflation is to construct a new nonlinear problem $H(u)$ with the same solutions as $G$, except for the $u^{\star}$ now known. We can then apply Newton's method to $H$, starting again from $u^0$, and (if it converges) it will converge to a different solution.
(In particular, with deflation we have only provided \emph{one} initial guess for the initial parameter value in our implementation.)
In this work, we construct the deflated problem $H$ via
\begin{equation*}
    H(u) \coloneqq \left(\frac{1}{\|u-u^{\star}\|^2} + 1\right)G(u) = 0,
\end{equation*}
where the norm used in this work is
\begin{equation*}
\|u\|^2 = \|(\delta \rho, Q)\|^2 = \int_\Omega (\delta \rho)^2 + \sum_i Q_i^2 \,\mathrm{d}x,
\end{equation*}
where $Q_i$ is the $i^\mathrm{th}$ component of the vector proxy for $Q$ (i.e., 2 components in two dimensions, 5 in three dimensions).
Under mild assumptions it can be proven that Newton's method applied to $H$ will not converge to $u^{\star}$ again.

With the idea of deflation for a single nonlinear problem at hand, we now briefly describe the deflated continuation algorithm, which is the combination of deflation
and continuation in the parameter value $\lambda$. Continuation is often used for difficult nonlinear problems to aid convergence.
The algorithm varies $\lambda$ in a specified grid of values $\lambda\in [\lambda_\mathrm{min},\lambda_\mathrm{max}]$.
Consider a continuation step from $\lambda^-$ to $\lambda^+$ and suppose that $m$ solutions $u_1^-, u_2^-, \dots, u_m^-$ are known at $\lambda^-$.
The step proceeds in two phases.
First, each
solution $u_i^-$ is continued from $\lambda^-$ to $\lambda^+$ to yield $u_i^+$ (using
arclength, tangent, secant or standard continuation~\cite{seydel2010}; in this work, we use standard continuation for Scenarios I and II, and secant continuation for the more difficult Scenario III). As each solution $u_i^+$ is computed, it is deflated
away from the nonlinear problem at $\lambda^+$.
Once all known solutions have been continued, the search phase of the algorithm begins.
Each previous solution $u_i^-$ is used again as an initial guess for the nonlinear problem at $\lambda^+$.
As discussed above, deflation ensures that the Newton iteration does not converge to any of the previously known solutions $u_i^+$ and, thus, if Newton's method converges it must converge to a new, unknown solution.
If an initial guess yields a new solution, it is deflated and the initial guess is used repeatedly until failure.
Once all initial guesses from $\lambda^-$ have been exhausted, this continuation step completes and the algorithm proceeds to the next step.
The above two phases are repeated for each continuation step until $\lambda$ reaches the desired target value $\lambda_\mathrm{max}$.

\subsection{Stability calculation}

To compute the stability of each solution profile, we calculate the inertia of the Hessian matrix of the energy functional with a Cholesky factorization,
implemented in MUMPS~\cite{mumps}.
If the Hessian matrix is positive semidefinite, we characterize the
solution as stable, while any nonzero number of negative eigenvalues
characterizes an unstable solution~\cite{num-op99}. No zero eigenvalues
of Hessians were observed in this work, i.e., the stable solutions all in fact had positive-definite Hessian matrices.

For a handful of parameter values where deflated continuation yields a solution of lowest energy that is unstable (i.e., does not find a candidate ground state), we then calculate the eigendirections of negative curvature using the Krylov--Schur algorithm~\cite{stewart2002} implemented in SLEPc~\cite{hernandez2005}. We then perturb the lowest-energy solution along its eigendirections of negative curvature and employ the bounded Newton line search algorithm of TAO~\cite{tao} to converge to a stable solution of minimal energy.

\section{Details of Each Simulation}
We give further details for the configuration of each example.

For the choice of parameters, we mainly use the values suggested in Pevnyi et al.~\cite{Pevnyi:2014kw},
occasionally varying them based on physical intuition (e.g., choosing a larger wave number $q$ to achieve thinner layers, or a larger anchoring weight $w$ to more strongly enforce the boundary conditions).
The new parameters that do not appear in the model of Pevnyi et al.~($l$ and $w$) were chosen via unreported initial numerical experiments.

\subsection{Scenario I: defect free}

This is a simple example proposed by the work of Williams and Kl\'eman \cite{10.1051/jphyscol:1975152} to examine the bending effect in smectics.
For a rectangle $\Omega = (-2,2)\times(0,2) $ with boundary labels
\begin{align*}
    &\Gamma_l = \left\{(x,y): x=-2\right\},&&\quad \Gamma_r = \left\{(x,y): x=2\right\},\\
    &\Gamma_b = \left\{(x,y): y=0\right\},&&\quad \Gamma_t = \left\{(x,y): y=2\right\},
\end{align*}
we strongly impose values of $Q$ consistent with $\mathbf{n}_e = (\cos\theta_0,-\sin\theta_0)$ at the bottom boundary, $\Gamma_b$, and with $\mathbf{n}_e=(\cos\theta_0,\sin\theta_0)$ at the top boundary, $\Gamma_t$, for fixed $\theta_0\in [0,\pi/2]$, that is to say,
\begin{equation*}
    \begin{aligned}
        Q &=
        \begin{bmatrix}
            (\cos\theta_0)^2-\frac{1}{2} & -\cos\theta_0\sin\theta_0 \\
            -\cos\theta_0\sin\theta_0 & (\sin\theta_0)^2-\frac{1}{2}
        \end{bmatrix} \quad && \text{on }\Gamma_b,\\
        Q &=
        \begin{bmatrix}
            (\cos\theta_0)^2-\frac{1}{2} & \phantom{-}\cos\theta_0\sin\theta_0\\
            \phantom{-}\cos\theta_0\sin\theta_0 & (\sin\theta_0)^2-\frac{1}{2}
        \end{bmatrix} \quad && \text{on }\Gamma_t,\\
    \end{aligned}
\end{equation*}
and impose periodic boundary conditions on the left and right boundaries, $\Gamma_l$ and $\Gamma_r$.
We discretize the domain $\Omega$ into $90\times 30$ quadrilateral elements.

The final form of the functional to be minimized in this scenario is
    \begin{align}
    F_\gamma(\delta\rho, Q)
    =\int_\Omega & \bigg( \frac{a}{2} \left(\delta\rho\right)^2 +\frac{b}{3} \left(\delta\rho\right)^3+\frac{c}{4} \left(\delta\rho\right)^4\nonumber\\
                 & + B\left| \mathcal{D}^2 \delta\rho + q^2 \left(Q+\frac{I_2}{2}\right) \delta\rho \right|^2\nonumber\\
                 & + \frac{K}{2}|\nabla Q|^2 -l \left(\text{tr}\left(Q^2\right)\right) + l \left(\text{tr}\left(Q^2\right)\right)^2 \bigg)\nonumber\\
        +& \sum_{e\in \mathcal{E}_I}\int_{e} \frac{1}{2h_e^3} \left( \llbracket \nabla \delta\rho \rrbracket \right)^2.
    \end{align}

We take the following initial guesses for $\delta\rho$ and $Q$:
\begin{equation}
    \label{eq:toriform}
    \delta\rho = 1, \quad Q=Q_0,
\end{equation}
where $Q_0=(\mathbf{n}_0\otimes\mathbf{n}_0-\frac{I_2}{2})$ with
\begin{equation*}
    \mathbf{n}_0 = \frac{1}{m}
        \begin{bmatrix}
             x\left(|x| - R\right)\\
             \left(|x|\right)y
        \end{bmatrix},
\end{equation*}
and
\begin{equation*}
    m=|x|\sqrt{\left(R-|x|\right)^2+y^2}.
\end{equation*}
Here, the initial guess for the $Q$-tensor is computed from a simplified two-dimensional mathematical representation of a family of tori, and we have taken the major radius $R=0.5$ in this implementation.

Finally, we specify the values of parameters in this experiment:
\begin{equation*}
    \begin{split}
        a = -10,~b =& 0,~c =10,~B = 10^{-5},~ K = 0.3,\\
                    & q=30 \text{ and } l = 30.
    \end{split}
\end{equation*}

\subsection{Scenario II: focal conic domains}

\begin{figure} 
\includegraphics{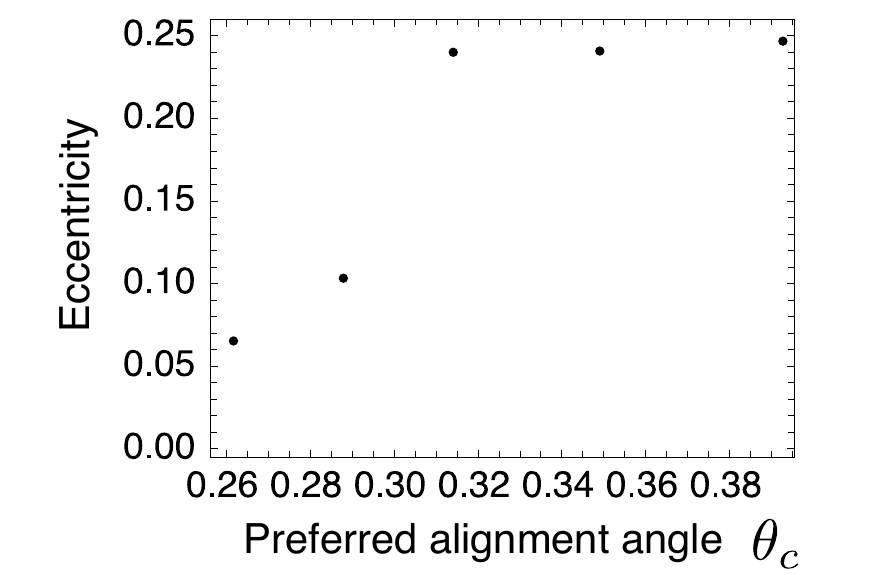}

\caption{\label{fig:Eccentricity}Eccentricity of FCD solutions as a function of preferred surface alignment angle.}

\end{figure}

We discretize the cuboid $\Omega=(-1.5,1.5)\times(-1.5,1.5)\times(0,2)$ into $6\times 6\times 5$ uniform hexahedra, to avoid a directional bias observed in numerical solutions with tetrahedra.
To simulate TFCDs or FCDs, we must impose boundary conditions (weakly or strongly) that respect their physical properties.  Thus,
we label the six boundary faces of $\Omega$ as
\begin{align*}
    &\Gamma_{left} = \{(x,y,z): x=-1.5\}, \, \Gamma_{right} = \{(x,y,z): x=1.5\},\\
    &\Gamma_{back} = \{(x,y,z): y=-1.5\}, \, \Gamma_{front} = \{(x,y,z): y=1.5\},\\
    &\Gamma_{bottom} = \{(x,y,z): z=0\}, \, \Gamma_{top} = \{(x,y,z): z=2\}.
\end{align*}

We consider the following surface energy
\begin{equation*}
    f_s(Q) = \int_{\Gamma_{bottom}} \frac{w}{2}\left| Q-Q_{radial}\right|^2
    + \int_{\Gamma_{top}} \frac{w}{2}\left| Q-Q_{vertical} \right|^2,
\end{equation*}
where $w$ denotes the weak anchoring weight,
\begin{equation*}
    Q_{radial} = \begin{bmatrix}
    \frac{x^2}{{x^2+y^2}}-\frac{1}{3} & \frac{xy}{{x^2+y^2}} & 0\\
    \frac{xy}{{x^2+y^2}} & \frac{y^2}{{x^2+y^2}}-\frac{1}{3} & 0\\
    0 & 0 & -\frac{1}{3}
\end{bmatrix}
\end{equation*}
represents an in-plane ($x$-$y$ plane) radial configuration of the director, and
\begin{equation*}
    Q_{vertical} = \begin{bmatrix}
    -\frac{1}{3} & 0 & 0\\
    0 & -\frac{1}{3} & 0\\
    0 & 0 & \frac{2}{3}
\end{bmatrix}
\end{equation*}
gives a vertical (i.e., along the $z$-axis) alignment configuration of the director.
Therefore, the final form of the functional to be minimized in the TFCD scenario is
\begin{align}
    F_\gamma&(\delta \rho, Q)
=\int_\Omega \bigg( \frac{a}{2} \left(\delta\rho\right)^2 +\frac{b}{3} \left(\delta\rho\right)^3+\frac{c}{4} \left(\delta\rho\right)^4 \nonumber\\
             & + B\left| \mathcal{D}^2 \delta\rho + q^2 \left(Q+\frac{I_3}{3}\right) \delta\rho \right|^2 \nonumber\\
             & +\frac{K}{2}|\nabla Q|^2 \nonumber\\
             &- \frac{l}{2} \left(\text{tr}(Q^2)\right) - \frac{l}{3} \left(\text{tr}(Q^3)\right) + \frac{l}{2} \left(\text{tr}(Q^2)\right)^2 \bigg) \nonumber\\
    + & \int_{\Gamma_{bottom}} \frac{w}{2}\left| Q-Q_{radial}\right|^2
    + \int_{\Gamma_{top}} \frac{w}{2}\left| Q-Q_{vertical} \right|^2 \label{eq:tfcd-energy}\\
    + & \sum_{e\in \mathcal{E}_I}\int_{e} \frac{1}{2h_e^3} \left( \llbracket \nabla \delta\rho \rrbracket \right)^2.\nonumber
\end{align}

For the FCD scenario, we only change the top boundary condition to perturb the preferred tilted director configuration.
We perturb the angle between the director and the $z$-axis on the top surface $\Gamma_{top}$ by $\theta_c$, thus adopting
\begin{equation*}
    Q_{c} =
    \begin{bmatrix}
        -\frac{1}{3} & 0 & 0\\
        0 & (\sin(\theta_c))^2-\frac{1}{3} & \sin(\theta_c)\cos(\theta_c)\\
        0 & \sin(\theta_c)\cos(\theta_c) & (\cos(\theta_c))^2-\frac{1}{3}
    \end{bmatrix}
\end{equation*}
instead of $Q_{vertical}$ in \eqref{eq:tfcd-energy}.
Note that when taking $\theta_c=0$, we return to the TFCD case.

Furthermore, we take the initial guesses:
\begin{equation*}
    \delta\rho = \cos(6\pi z), \quad Q=Q_{ic},
\end{equation*}
where $Q_{ic}=\left(\mathbf{n}_{ic}\otimes\mathbf{n}_{ic}-\frac{I_3}{3}\right)$ with
\begin{equation*}
    \mathbf{n}_{ic} = \frac{1}{m}
        \begin{bmatrix}
             x\left(\sqrt{x^2+y^2} - R\right)\\
             y\left(\sqrt{x^2+y^2}-R\right)\\
             z\left(\sqrt{x^2+y^2}\right)
        \end{bmatrix},
\end{equation*}
and
\begin{equation*}
    m=\sqrt{x^2+y^2}\sqrt{\left(R-\sqrt{x^2+y^2}\right)^2+z^2}.
\end{equation*}
Here, the initial guess for the $Q$-tensor is computed from the mathematical representation for a family of tori, and we have taken a major radius $R=1.5$ in our implementation.

We specify the values of parameters used in the (T)FCD experiments:
\begin{equation*}
    \begin{split}
        a = -10,~ b = 0,&~ c =10,~ B = 10^{-3},~ K = 0.03,\\
        q=10,&~ l = 30~ \text{and } w=10.
    \end{split}
\end{equation*}

As the Dupin cyclide has a confocal pair of a hyperbola and an ellipse, we fit a hyperbola to each solution with least squares (data points extracted via ParaView~\cite{ahrens2005}) and calculate its eccentricity (e.g., for a hyperbola expressed as $\frac{y^2}{a_{fit}^2}-\frac{z^2}{b_{fit}^2}=1$, its eccentricity is defined as $\frac{\sqrt{a_{fit}^2+b_{fit}^2}}{a_{fit}}$).
Then the eccentricity of the ellipse is the inverse of that of the confocal hyperbola. Values of eccentricity fitted from the solution set are shown as a function of the preferred surface alignment angle $\theta_c$ in Fig.~\ref{fig:Eccentricity} of this supplemental material. 

\subsection{Scenario III: oily streaks}

Let $L/\tau$ denote the aspect ratio of a rectangle $\Omega=(-L/\tau, L/\tau)\times (0,2)$.
To simulate oily streaks in the rectangle, $\Omega$, we weakly impose a planar (i.e., horizontal) boundary condition on the bottom face and a homeotropic (i.e., vertical) condition on the top surface.
We label the different boundaries of the rectangle as
\begin{align*}
    &\Gamma_l = \left\{(x,y): x=-L/\tau\right\},&&\quad \Gamma_r = \left\{(x,y): x=L/\tau\right\},\\
    &\Gamma_b = \left\{(x,y): y=0\right\},&&\quad \Gamma_t = \left\{(x,y): y=2\right\}.
\end{align*}
Then the following surface energy is imposed:
\begin{equation*}
    f_s(Q) = \int_{\Gamma_b} \frac{w}{2}\left| Q-Q_{bottom}\right|^2
    + \int_{\Gamma_t\cup\Gamma_l\cup\Gamma_r} \frac{w}{2}\left| Q-Q_{top} \right|^2,
\end{equation*}
where $w$ is the weak anchoring weight and two weakly prescribed configurations $Q_{bottom}$ and $Q_{top}$ are given by
\begin{equation*}
    Q_{bottom} = \begin{bmatrix}
    \frac{1}{2} & 0\\
    0 & -\frac{1}{2}
\end{bmatrix},
\end{equation*}
yielding horizontally aligned directors, and
\begin{equation*}
    Q_{top} = \begin{bmatrix}
    -\frac{1}{2} & 0\\
    0 & \frac{1}{2}
\end{bmatrix},
\end{equation*}
yielding vertically aligned directors.

In the aspect-ratio continuation experiment, we always discretize the domain $\Omega$ into $90\times 30$ quadrilateral elements, even as we change the domain size by varying the aspect ratio, $L/\tau$.
The final form of the functional to be minimized in this scenario is
    \begin{align}
    F_\gamma(\delta\rho, Q)
    =\int_\Omega & \bigg( \frac{a}{2} \left(\delta\rho\right)^2 +\frac{b}{3} \left(\delta\rho\right)^3+\frac{c}{4} \left(\delta\rho\right)^4 \nonumber \\
                 & + B\left| \mathcal{D}^2 \delta\rho + q^2 \left(Q+\frac{I_2}{2}\right) \delta\rho \right|^2 \nonumber \\
                 & + \frac{K}{2}|\nabla Q|^2 -l \left(\text{tr}\left(Q^2\right)\right) + l \left(\text{tr}\left(Q^2\right)\right)^2 \bigg) \nonumber \\
        + & \int_{\Gamma_{b}} \frac{w}{2}\left| Q-Q_{bottom}\right|^2 \nonumber\\
    + & \int_{\Gamma_{t}\cup\Gamma_l\cup\Gamma_r} \frac{w}{2}\left| Q-Q_{top} \right|^2 \nonumber\\
        + & \sum_{e\in \mathcal{E}_I}\int_{e} \frac{1}{2h_e^3} \left( \llbracket \nabla \delta\rho \rrbracket \right)^2.
    \end{align}
    In the bifurcation diagram (see Fig.~3D in the main text) we calculate the energy per unit area to identify which aspect ratio is energetically favorable.

    We take the same form of the initial guess for $\delta\rho$ and $Q$ as in \eqref{eq:toriform} but with a larger major radius $R=1$.

Finally, we specify the values of parameters in this experiment:
\begin{equation*}
    \begin{split}
        a = -10,~ b = 0,&~ c =10,~ B = 10^{-5},~ K = 0.3,\\
        q=30,&~ l = 1~ \text{and } w=10.
    \end{split}
\end{equation*}

\section{Further numerical results in scenario I}

\begin{figure*}
    \includegraphics[scale=0.55]{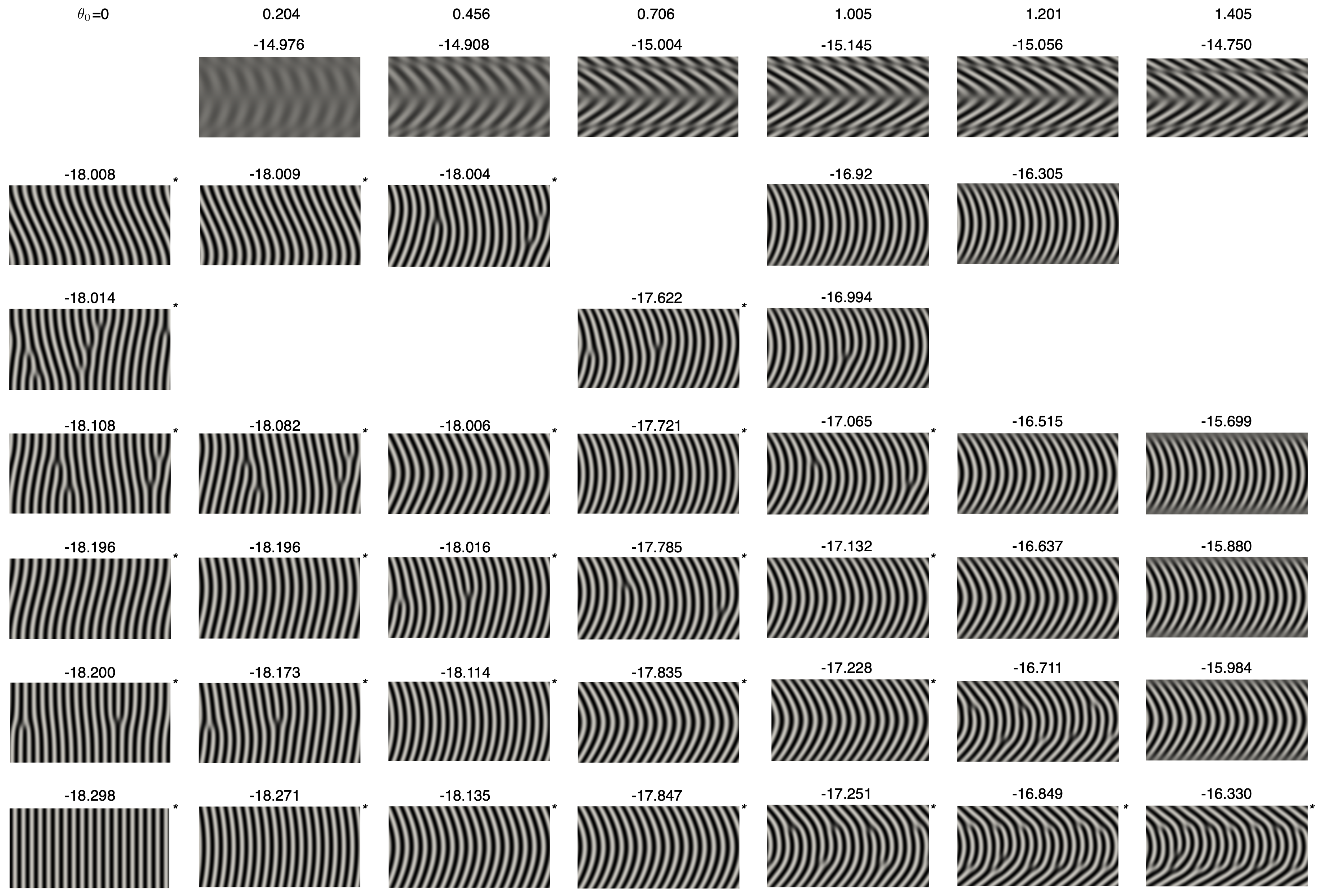}

\caption{\textbf{\label{fig:figwall}Stationary states in the defect-free scenario.}
Stationary states obtained at different values of $\theta_0$.
For each solution, the value of the energy functional is displayed above it and we specify the stable profiles with asterisks.
The bottom row depicts the lowest-energy solution found for each value of $\theta_0$.}

\end{figure*}

Here, we present more solutions found using the {deflation} technique in our implementation of the defect-free experiment.
In Fig.~\ref{fig:figwall} of this supplemental material, we show selected stationary states as a function of $\theta_0$.
For each state, we display the value of the energy functional
    \begin{align*}
    F(\delta\rho, Q)
    =\int_\Omega & \bigg( \frac{a}{2} \left(\delta\rho\right)^2 +\frac{b}{3} \left(\delta\rho\right)^3+\frac{c}{4} \left(\delta\rho\right)^4\\
                 & + B\left| \mathcal{D}^2 \delta\rho + q^2 \left(Q+\frac{I_2}{2}\right) \delta\rho \right|^2\\
     & + \frac{K}{2}|\nabla Q|^2 -l \left(\text{tr}\left(Q^2\right)\right) + l \left(\text{tr}\left(Q^2\right)\right)^2 \bigg)
    \end{align*}
    per unit area.
    For each column (i.e., fixed value of $\theta_0$), we organize the stationary states in an energy-decreasing order and identify stable profiles with asterisks.
The bottom row depicts the lowest-energy minimizers found, all of which are stable.

\section{Descriptions of attached videos}

For better illustration of the numerical results, we have included three videos in this supplemental material. We give descriptions of these videos here.

The video \emph{scenario-i-lowest-energy-in-theta-zero.mp4} presents the stationary configurations of lowest energy found in the defect-free experiment described in Scenario I as we vary the applied bend deformation $\theta_0\in [0,\pi/2]$.
The profiles shown in the video are all stable.

The video \emph{scenario-ii-pi12.mp4} illustrates the three (stable) solutions found (FCD, single and double screw dislocations) in the implementation of Scenario II at $\theta_c=\pi/12$.
It depicts the zero-isosurfaces of the smectic density variation field $\delta \rho$. The isosurfaces are colored by height (the $z$-coordinate) to assist in depth perception. The problem solved is stationary; the time axis of the video is used to illustrate the internal structure of the layers.

The video \emph{scenario-iii-lowest-energy-in-r.mp4} depicts the lowest-energy configurations discovered as we vary the aspect ratio $L/\tau \in [1, 5]$ for the oily streaks example in Scenario III.
The presented profiles are all stable.

\section{Code availability}

For reproducibility, both the solver code \cite{zenodo-smectica} and the exact version of Firedrake used \cite{zenodo/firedrake-smecticA} to produce the numerical results of this paper have been archived on Zenodo. An installation of Firedrake with components matching those used in this paper can be obtained by following the instructions at \url{https://www.firedrakeproject.org/download.html} with

\begin{small}
\begin{verbatim}
python3 firedrake-install --doi 10.5281/zenodo.4441123
\end{verbatim}
\end{small}

Defcon version \#11e883c should then be installed, as described in \url{https://bitbucket.org/pefarrell/defcon/}.

\bibliographystyle{apsrev4-2}
\bibliography{references}